\documentclass[a4paper,10pt,twoside]{cpc-hepnp}

\usepackage{multicol}
\usepackage{graphicx}
\usepackage{booktabs}
\usepackage{amssymb,bm,mathrsfs,bbm,amscd}
\usepackage[tbtags]{amsmath}
\usepackage{lastpage}
\usepackage{CJK}

\begin{document}

\fancyhead[c]{\small Submitted to Chinese Physics C} \fancyfoot[C]{\small \thepage}


\title{Multi-wavelength study of the MGRO J2019+37}

\author{%
      HOU Chao$^{1;1)}$\email{houchao@ihep.ac.cn}%
\quad CHEN Song-Zhan$^{1}$
\quad YUAN Qiang$^{1}$\\
\quad CAO Zhen$^{1}$
\quad HE Hui-Hai$^{1}$
\quad SHENG Xiang-Dong$^{1}$
}
\maketitle

\address{%
$^1$ Institute of High Energy Physics, Chinese Academy of Sciences, Beijing 100049, China\\
}

\begin{abstract}
MGRO J2019+37 within the Cygnus region is a bright and extended source  revealed by Milagro  at 12-35 TeV.
This source is almost as bright as Crab Nebula in northern sky, while it is not confirmed by ARGO-YBJ around TeV.
Up to now, no obvious counterpart at low energy wavelengths has been found.
Hence, MGRO J2019+37 becomes mysterious and its VHE $\gamma$-ray emission mechanism is attractive.
In this paper, a brief summary of the multi-wavelength observations from Radio to $\gamma$-ray is presented.
All the available data from the XMM-Newton and INTEGRAL at X-ray, and the \emph{Fermi}-LAT at $\gamma$-ray bands were used to get constraint on its emission flux at low
energy wavelengths.
Then, its possible counterparts and the VHE emission mechanism are discussed.
\end{abstract}



\footnotetext[0]{\hspace*{-3mm}\raisebox{0.3ex}{$\scriptstyle\copyright$}2013
Chinese Physical Society and the Institute of High Energy Physics
of the Chinese Academy of Sciences and the Institute
of Modern Physics of the Chinese Academy of Sciences and IOP Publishing Ltd}%

\begin{multicols}{2}

\section{Introduction}
Cygnus region is an active massive star formation and destruction portion on the Galactic plane, with coordinate range of (l$\in[65^{\circ},85^{\circ}]$,
b$\in[-2^{\circ},+2^{\circ}]$) and distance of 1$-$2 kpc from us.
It is the brightest diffuse $\gamma$-ray source in the northern sky as revealed by \emph{Fermi}-LAT at GeV, ARGO-YBJ at TeV and Milagro at 15 TeV \cite{FermiLAT2FGL,milagro15TeV,mallicrc2011}.
As containing a great deal of molecular clouds and being rich in potential cosmic-ray acceleration sites, such as Wolf-Rayet stars, OB associations andsupernova remnants, the Cygnus region becomes a hot land for scientists to study the origin of cosmic ray \cite{argocygregion,bi2009}.

MGRO J2019+37 detected by Milagro within the Cygnus region is towards the Cyg OB1 association.
It is the brightest source of the three new extended sources discovered by Miagro experiment when it surveyed the northern Galactic plane\cite{milagro2007}.
The extension   is $\sigma$= $0.32^{\circ}\pm0.12^{\circ}$ for a symmetric two-dimensional Gaussian shape.
Its measured flux is about 80\%   Crab unit at 20 TeV \cite{milagro2007survey}.
This source is suspected to be associated with the GeV pulsar J2021+3651 \cite{milagro2009fermi}.
About 0.9$^{\circ}$ away from MGRO J2019+37, Tibet AS$\gamma$ collaboration reported a preliminary 5.8$\sigma$ excess in \cite{Tibet2007}. While only marginal signal was reported in their later formal result \cite{Tibet2008,Tibet2010}.

The ARGO-YBJ experiment is a full coverage extensive air shower (EAS) array with a large field of view (FOV) at a high altitude of 4300 m.
The threshold is around 300 GeV, which is much lower than any previous EAS arrays. The other two bright extended sources discovered by Milagro, i.e., MGRO J2031+41 and MGRO J1908+06, have been confirmed by ARGO-YBJ with  significance greater than 5$\sigma$\cite{argocygregion,argoJ1908+06,argosurvey2013}.
MGRO J2031+41 is also located in Cygnus region just nearby MGRO J2019+37.
The energy spectra of these two sources measured by ARGO-YBJ are consistent with that measured by Milagro.
Besides these, ARGO-YBJ detected another extended source HESS J1841$-$055 \cite{argoHESSJ1841}.
Unexpectedly, ARGO-YBJ detected little signal from the brightest Milagro  source MGRO J2019+37,
and the derived flux upper limits at the 90\% confidence level (c.l.) are lower than the Milagro flux at energies below 5 TeV \cite{argocygregion}.
The ARGO-YBJ upper limits do not conflict with the Milagro $1\sigma$ error region in a new analysis applied to the Milagro data from 2005 to 2008
\cite{milagro2012}, while they constrain the flux should be lower than the best-fitting value derived by Milagro.
In such a situation, a peak structure is formed with the energy as high as about 10 TeV. 

VERITAS is a narrow FOV imaging atmospheric Cherenkov telescope with excellent energy and angular resolution  ranging  from hundreds of GeV to multi TeV.
In 2007, VERITAS had surveyed the Cygnus region with a sensitivity of 6.3\% Crab unit, but no
emission from MGRO J2019+37 was detected \cite{VERITAS2009}.
In 2010, with further deep observations   better than  1\% Crab unit, VERITAS revealed a complex TeV emitting structure at the position of MGRO
J2019+37 which is likely powered by multiple sources \cite{VERITAS2011}.

The MGRO J2019+37 region was surveyed by the Giant Metrewave Radio Telescope (GMRT)  at the frequency of 610 MHz and the 3.5 m telescope in Calar Alto at
the near-infrared $K_s$-band.
A catalogue of 362 radio sources and $\sim3\times10^5$ near-infrared sources were detected \cite{radioandinfrared}.
Some peculiar sources are noticeable, such as the pulsar PSR J2021+3651, two new radio-jet sources, the radio source NVSS J202032+363158 and the
H\uppercase\expandafter{\romannumeral2} region Sh 2-104 containing two star clusters.

The MGRO J2019+37 region was also observed by XMM-Newton  at 1$\sim$8 keV   with a high sensitivity.
A $\sim$20' extended  emission around PSR J2021+3651 and an UCH\uppercase\expandafter{\romannumeral2} region in Sh 2-104 were detected.
In the GeV band, EGRET, \emph{Fermi}-LAT and AGILE  all detected some point sources with this extended region.
Among them, PSR J2021+3651 detected by $Fermi$-LAT is a spin-powered radio pulsar whose spectrum has a cutoff at 10 GeV.

To sum up, no similar morphology as that of TeV emission is  found at radio, optical, X-ray and GeV $\gamma$-ray bands. Therefore, no definite counterpart of MGRO J2019+37 is found at low energy bands.

The existing observation by the low energy bands telescopes doesn't give a flux constraint especially for the MGRO J2019+37 extended region.
In order to better understanding the TeV emission mechanism, a multi-wavelength observation, especially at X-ray and $\gamma$-ray bands,   is quite
necessary.
For such a large extended region,  observations for MGRO J2019+37  had better to be implemented by  detectors with wide  FOV.
In this paper, all the available data from the INTEGRAL at hard X-ray and the \emph{Fermi}-LAT at  $\gamma$-ray bands, which are wide FOV, are used to   constrain the  emission flux from MGRO J2019+37.
In addition, the soft X-ray data from the narrow FOV detector XMM-Newton are also analyzed.
Then, its possible origins and the corresponding  emission mechanism are discussed.

\section{Multi-wavelength analysis}
Since the angular resolution of ground based particle detector is relatively
poor, it is difficult to identify the low energy counterpart of a TeV
$\gamma$-ray source given there are usually more than one low energy sources
in the error box of the TeV source. Variability is a good identifier to
find the counterpart for variable sources. However, it becomes more
difficult for a steady source. In this case, we may need a more detailed
study of the properties of all the observed sources, such as spectrum,
flux, pulsation and even polarization to explore their possible connection
with the target source. In the following, we try to find the low
energy counterpart of MGRO J2019+37 and discuss its multi-wavelength
emission mechanism.

\subsection{\emph{Fermi}-LAT GeV gamma-ray}
\emph{Fermi}-LAT is an imaging high-energy $\gamma$-ray telescope covering
the energy band from 20 MeV to 300 GeV. The angular resolution is about
$3.5^{\circ}$ at 100 MeV, improved to about $0.1^{\circ}$ at 10 GeV
\cite{fermiRXJ1713}. The FOV of \emph{Fermi}-LAT covers about 20\% of the
sky at a time. It scans continuously, and covers the whole sky every
three hours. The LAT data from a region of interest (ROI) centered on
MGRO J2019+37 ($304.63^{\circ},36.88^{\circ}$) with a radius of
$20^{\circ}$ were downloaded from Fermi Science Support
Center\footnote{http://fermi.gsfc.nasa.gov/ssc/}. The observational time
is from 4 August 2008 to 4 August 2012.

The data analysis was performed following the standard procedure with a
binned maximum-likelihood method. The model adopted in the likelihood
fitting included the diffuse backgrounds with both the Galactic and
isotropic components, as well as the point sources in 2FGL catalog
\cite{FermiLAT2FGL}. Within $1^{\circ}$ region of MGRO J2019+37, there are three
point sources in 2FGL catalog, which are 2FGL J2021+3651, 2GFL
J2018+3626 and 2FGL J2015.6+3709. 2FGL J2021+3651 is identified as
a pulsar, and 2FGL J2015.6+3709 is identified as an active galactic
nucleus (AGN). 2GFL J2018+3626 is unidentified, and it may be
the counterpart of MGRO J2019+37.

The left panel of Fig.~\ref{fermi_wide_fig} shows the residual
$\gamma$-ray counts map of the sky around MGRO J2019+37, after subtracting
the diffuse backgrounds. To reduce the influence of pulsar PSR J2021+3651,
which has a spectral cutoff at about 10 GeV \cite{fermi2009}, we adopt
an energy threshold of 10 GeV to show the counts map and for the
spectral analysis below. The counts map is smoothed with a $0.3^{\circ}$
width Gaussian kernel. From the counts map we see that 2GFL J2018+3626
has few photons above $10$ GeV. The spectrum given in 2FGL catalog of this
source is also very soft at high energies. Since other two sources located
in the vicinity of MGRO J2019+37 have been identified as a pulsar and an AGN,
we think that none of these three sources will serve as the counterpart
of MGRO J2019+37. The right panel of Fig.~\ref{fermi_wide_fig} gives the
residual map after further subtracting the known 2FGL sources. It shows
that there is not any excess at the location of MGRO J2019+37.

\end{multicols}
\begin{center}
\includegraphics[width=0.43\textwidth]{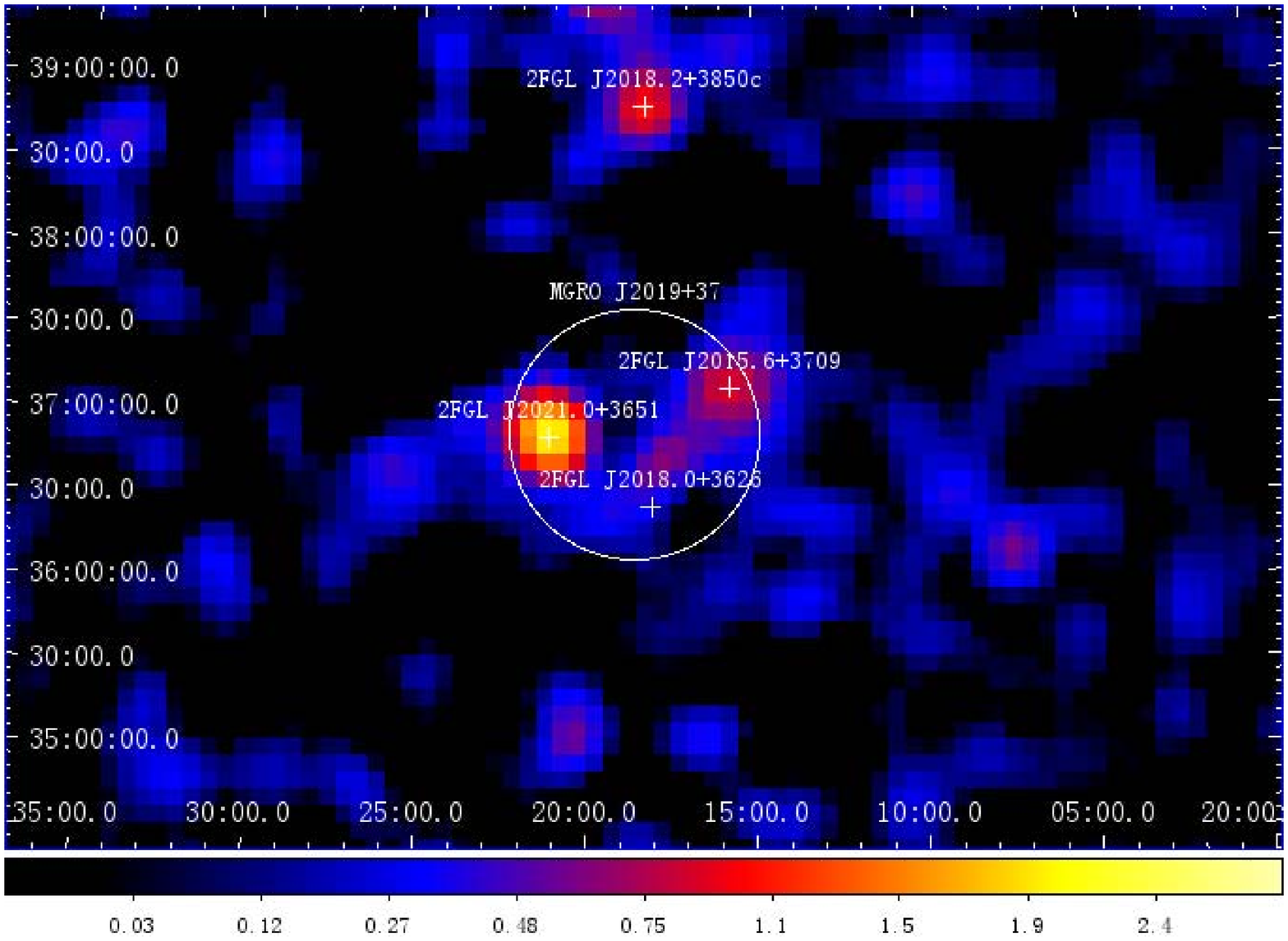}
\includegraphics[width=0.43\textwidth]{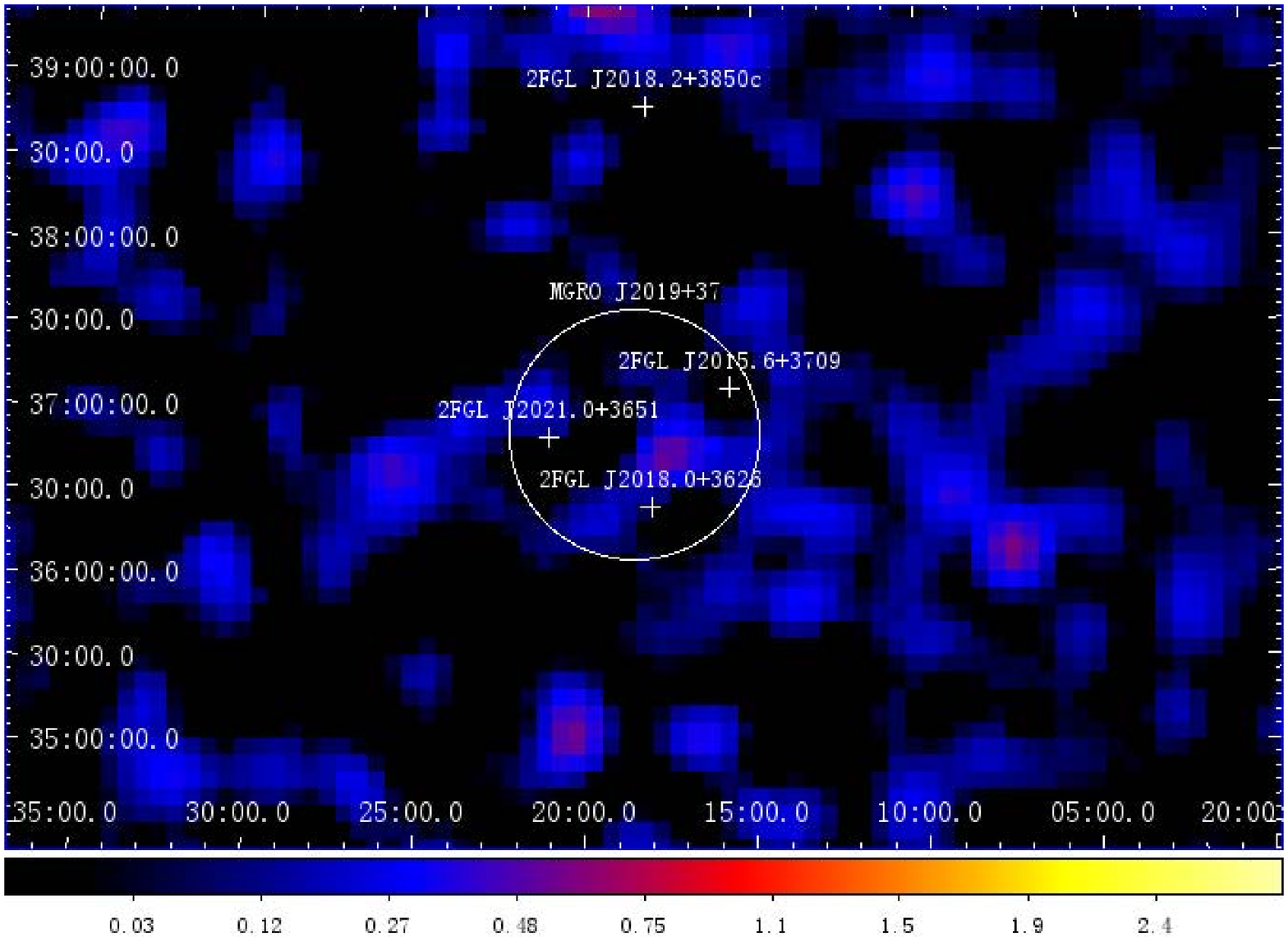}
\figcaption{\label{fermi_wide_fig} Residual counts map of $\gamma$-ray photons above 10 GeV using the data observed
by \emph{Fermi}-LAT, after subtracting the diffuse backgrounds (left) and the
diffuse backgrounds together with the known 2FGL sources (right).
The circle shows the target region of MGRO J2019+37 with diameter of
$1^{\circ}$, and the crosses label the known sources in LAT 2FGL catalog. (Color online)}
\end{center}
\begin{multicols}{2}

\end{multicols}
\begin{center}
\includegraphics[width=0.43\textwidth]{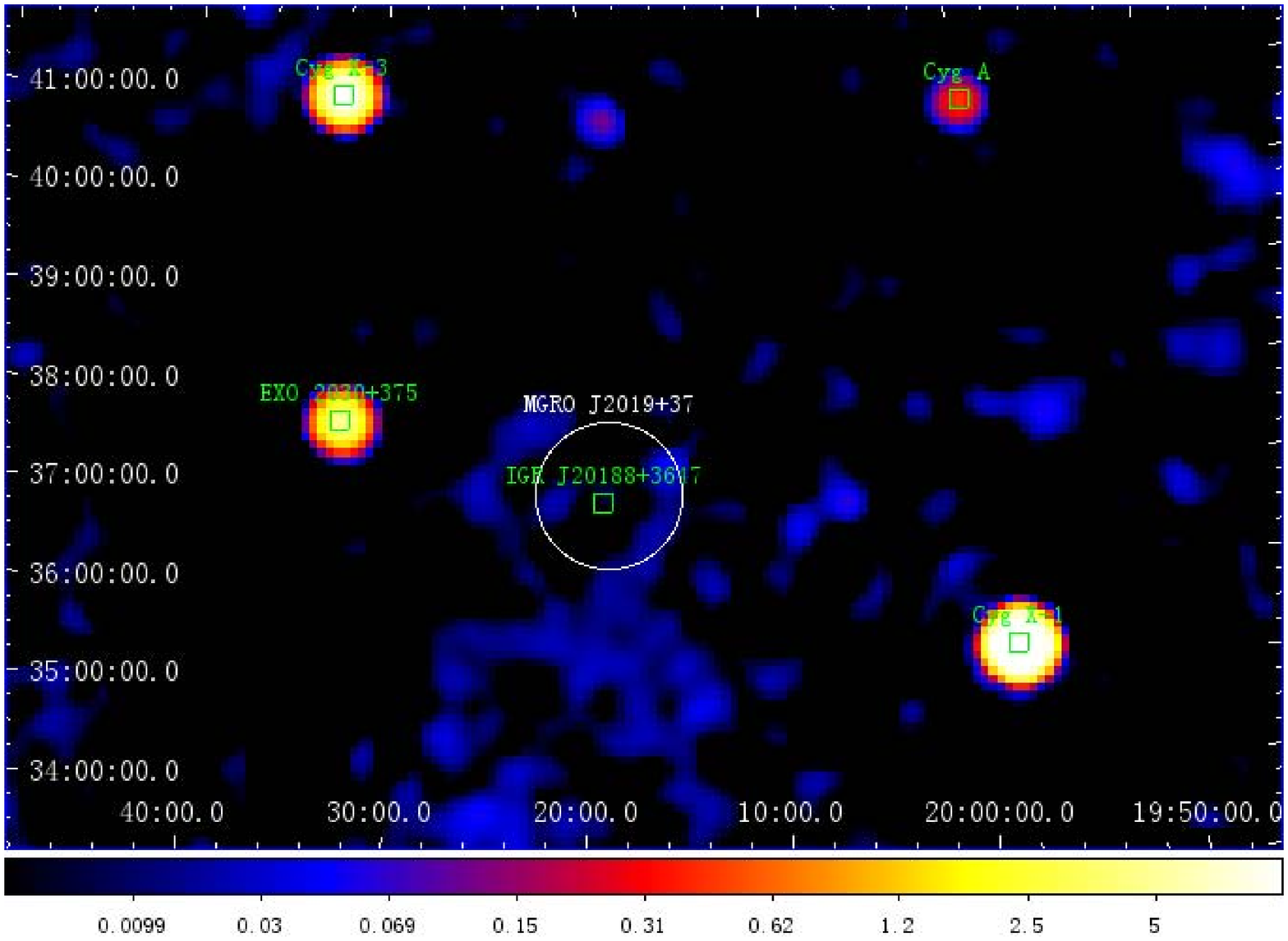}
\includegraphics[width=0.43\textwidth]{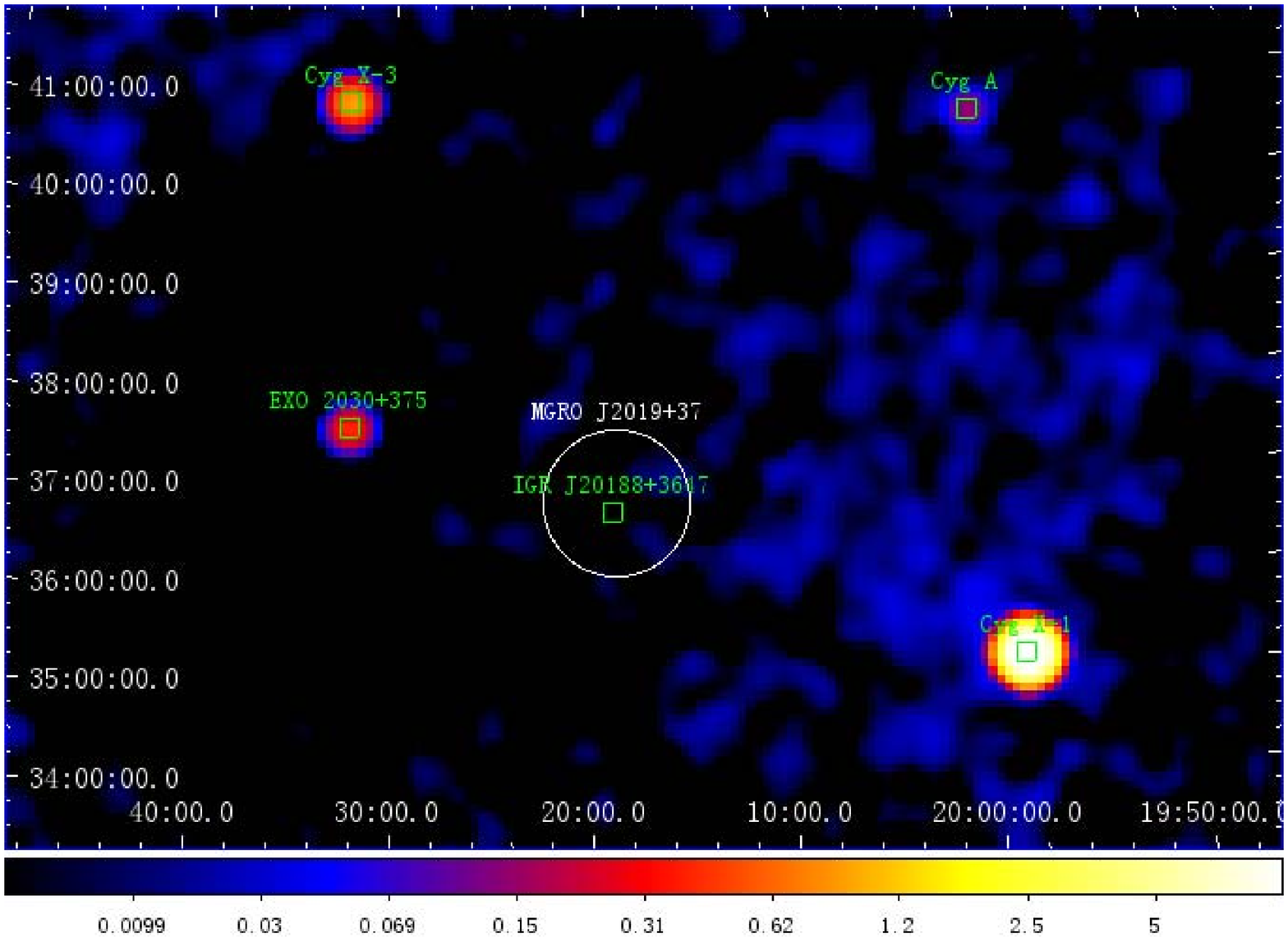}
\figcaption{\label{INTEGRAL_image_simp_fig}Count rate map of X-ray photons using the data observed by INTEGRAL IBIS/ISGRI
from 2002 to 2012, smoothed with a $0.25^{\circ}$ Gaussian kernel. The left
panel is for 20$-$60 keV, and the right panel is for 60$-$200 keV.
The circle shows the target region of MGRO J2019+37 with diameter of
$1^{\circ}$, and the squares represent the postions of known sources in
the INTEGRAL catalog. The point source located in MGRO J2019+37 circle,
IGR J20188+3647, is identified as a transient point source. (Color online)}
\end{center}
\begin{multicols}{2}

Then we add MGRO J2019+37 in the model and re-do the likelihood analysis.
MGRO J2019+37 is modeled with a two-dimensional Gaussian template with
$\sigma=0.35^{\circ}$ as revealed in TeV observation \cite{milagro2007}.
No strong signal with significance $>2\sigma$ is found. Therefore we derive
the upper limits of MGRO J2019+37 with \emph{Fermi}-LAT data.
Assuming a power-law index of $2$, we find the 95\% c.l. upper
limits of the fluxes are $2.1\times10^{-10}$ cm$^{-2}$s$^{-1}$ for $10-31$ GeV,
$8.0\times10^{-11}$ cm$^{-2}$s$^{-1}$ for $31.1-97$ GeV,
and $5.0\times10^{-11}$ cm$^{-2}$s$^{-1}$ for $97-300$ GeV, respectively.

\subsection{INTEGRAL IBIS/ISGRI hard X-ray}
IBIS (Imager on Board the INTEGRAL Satellite) is one of the
instruments of the INTEGRAL X-ray telescope. It works in the energy
range from $\sim$15 keV to several MeV. The FOV of IBIS is
$8.33^{\circ}\times8.00^{\circ}$ ($19^{\circ}\times19^{\circ}$) for fully
(50\%) coded mode, and the angular resolution (full width half maximum,
FWHM) is $0.2^{\circ}$. There are two detectors:
the Integral Soft $\gamma$-Ray Imager (ISGRI), which is a semi-conductor
array optimized for lower energies (18 keV-1 MeV), and the PIxelated
Ceasium Iodide (CsI) Telescope (PICsIT), which is a crystal scintillator
sensitive for higher energies (175 keV-10 MeV) \cite{INTEGRAum}.

The data and standard analysis software OSA10.0 are downloaded from the
INTEGRAL official website\footnote{http://www.isdc.unige.ch/integral/archive}.
In this analysis we adopt the ISGRI data recorded from 2002 to 2012 for
all observing numbers. Due to the quality of achieved data,
we restrict the analysis in the energy range $20-200$ keV, and divide the data
into two bands, 20$-$60 keV and 60$-$200 keV respectively.
Fig.~\ref{INTEGRAL_image_simp_fig} shows the count rate maps in these
two energy bands. These maps are smoothed with a 0.25$^{\circ}$ width
Gaussian kernel. The circle shows MGRO J2019+37 region with diameter
of $1^{\circ}$, and the white squares shows the positions of the sources discovered by INTEGRAL.
IGR J20188+3647, which is identified as a supergiant
fast X-ray transient (SFXT) source, shows a fast rise (10 minutes) followed
by a slow decay (50 minutes) \cite{J20188+3647}, and thus may be not the
counterpart of the extended TeV source MGRO J2019+37. According to these
count rate maps, there is no significant excess coincident with MGRO
J2019+37. We will estimate the upper limits of MGRO J2019+37 in the hard
X-ray band.

A circle with radius of 0.54$^{\circ}$ centered on MGRO
J2019+37, which encloses 68\% of the events from MGRO J2019+37 extended area taking into account of the point spread function (PSF) of ISGRI,
is used to calculate the count rate.
The expected background count rate is estimated from six
other circular regions with the same radius but 1.28$^{\circ}$ away from
the center of MGRO J2019+37. We use Helene method \cite{Helene} to
calculate the 95\% c.l. upper limit of the count rate. Since there is no
spectral analysis script for extended source analysis in OSA10.0 software,
we adopt a method proposed by Swift-BAT collaboration \cite{swift2010} to
derive the flux of an extended source through comparing the count rate with
that of the standard candle Crab Nebula. The source flux can be calculated by
\begin{equation}
{\rm source~flux = \frac{source~count~rate}{Crab~count~rate}\times Crab~flux}
 \end{equation}
in each energy band. To test this method, we apply it to the point sources
Cyg X-1 and Cyg X-3, which are close to MGRO J2019+37 as shown in Fig.~\ref{INTEGRAL_image_simp_fig}.
We compare the fluxes derived by this method
with that derived by the standard method for point source analysis conducted
by the OSA10.0. Although the spectral shapes
of Cyg X-1 and Cyg X-3 are very different from that of Crab Nebula, the
fluxes obtained by the two methods are consistent with each other within
10\%. To estimate the flux upper limit from MGRO J2019+37,
a power-law spectrum with index of 2.0 is assumed in this work, which is
very closed to the spectrum of Crab Nebula. Therefore, the systematic error
is expected to be smaller than 10\%. The final 95\% c.l. upper limits are
$1.4\times10^{-4}$ cm$^{-2}$s$^{-1}$ for $20-60$ keV and
$6.0\times10^{-5}$ cm$^{-2}$s$^{-1}$ for $60-200$ keV.

\subsection{XMM-Newton soft X-ray}

XMM-Newton has observed MGRO J2019+37 extended region with the European
Photon Imaging Camera (EPIC). EPIC has one pn and two MOS cameras, which
covers the energies from 0.2 to 12 keV with an energy resolution of
0.15 keV at 1 keV. Their FOV is 30', and the on-axis resolution angle is
about 6'' (FWHM) and 15'' (half-power diameter).

There are four archival observations of the MGRO J2019+37 extended region.
Two of the observations focus on the pulsar wind nebula (PWN) G75.2+0.1.
The third one points to IGR J20188+3647 and the fourth points to MGRO
J2019+37. The pulsar PSR J2021+3651 and its PWN G75.2+0.1, HII regions
sh2-104 and WR141 have been detected by EPIC in this extended region.
The mosaic image of this region can be found in Fig.~\ref{XMM}, which is  presented in \cite{XMM2010}.
There are no other candidate sources in $1^{\circ}$ region of MGRO J2019+37.
The PWN and HII region might be VHE emitters. Therefore we choose PWN
G75.2+0.1 and HII region sh2-104 for spectral analysis.


\begin{center}
\includegraphics[width=0.45\textwidth]{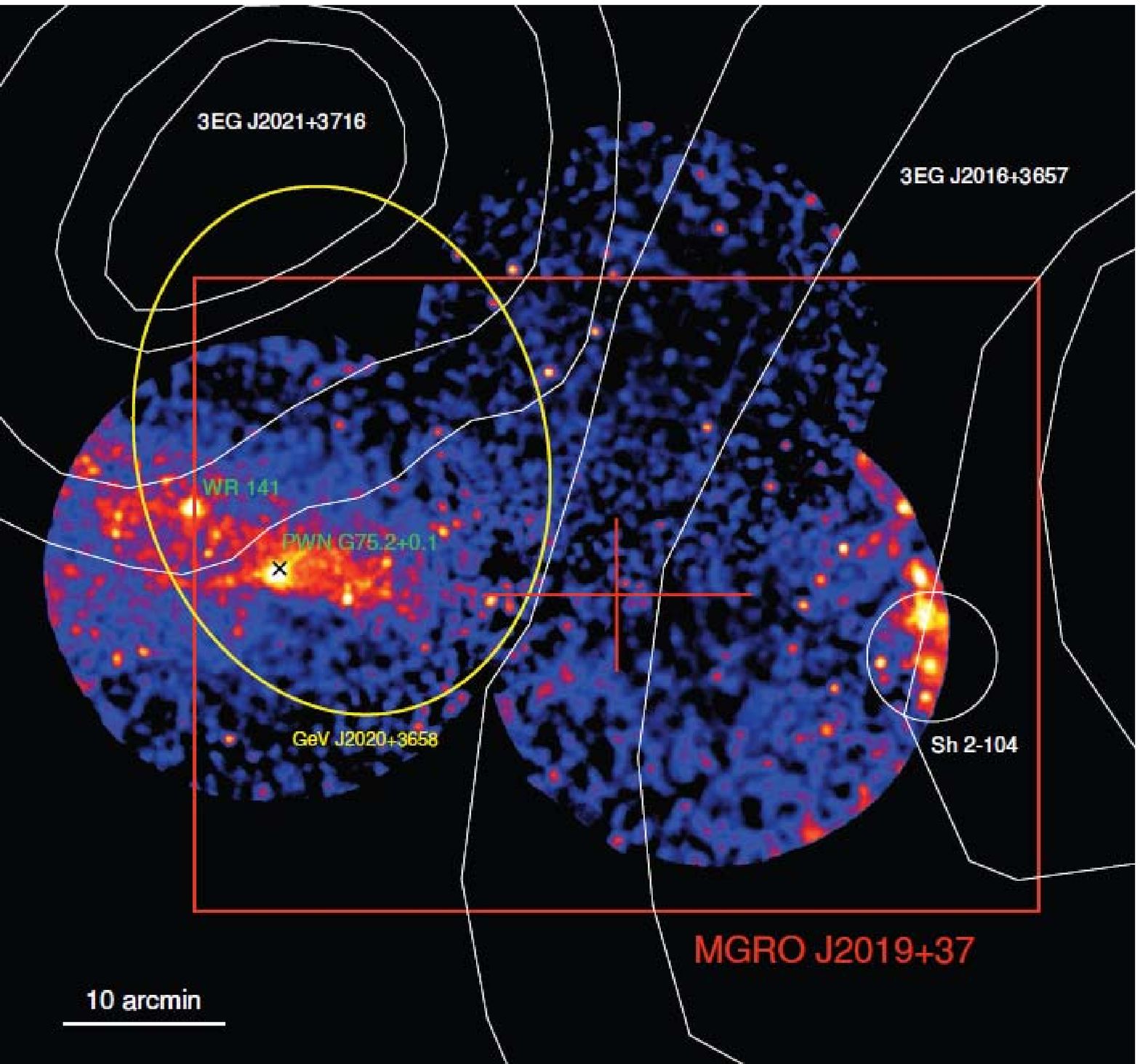}
\figcaption{\label{XMM} XMM-Newton background subtracted and exposure corrected X-ray (1keV-8keV) mosaic image of the MGRO J2019+37 region, which is  presented in \cite{XMM2010}. The central cross and box indicate the Gravity Center and its positional uncertainty including statistic and systematic errors of the TeV emission from MGRO J2019+37 \cite{milagro2007}. The black cross indicates the position of PWN G75.2+0.1. This picture also label the HII region Sh 2-104 and the Wolf Rayet star WR 141. North is up and East is left. (Color online)}
\end{center}

\end{multicols}
\begin{table}[t]
\begin{center}
\tabcaption{ \label{Fit-results}  The spectral fitting results obtained using XMM-Newton data.}
\footnotesize
\begin{tabular*}{170mm}{@{\extracolsep{\fill}}ccccc}
\toprule source name & model & $n_{\rm H}$/($4\times10^{22}$ cm$^{-2}$) & index & flux/($2-10$ keV, erg cm$^{-2}$s$^{-1}$) \\
\hline
PWN G75.2+0.1& absorbed power-law&0.31&1.44$_{-0.17}^{+0.18}$ & $2.62\times10^{-12}$\\
sh2-104& absorbed power-law&2.70&2.09$_{-0.73}^{+0.88}$& $4.54\times10^{-13}$ \\
\bottomrule
\end{tabular*}%
\end{center}
\end{table}

\begin{multicols}{2}

\begin{center}
\includegraphics[width=0.5\textwidth]{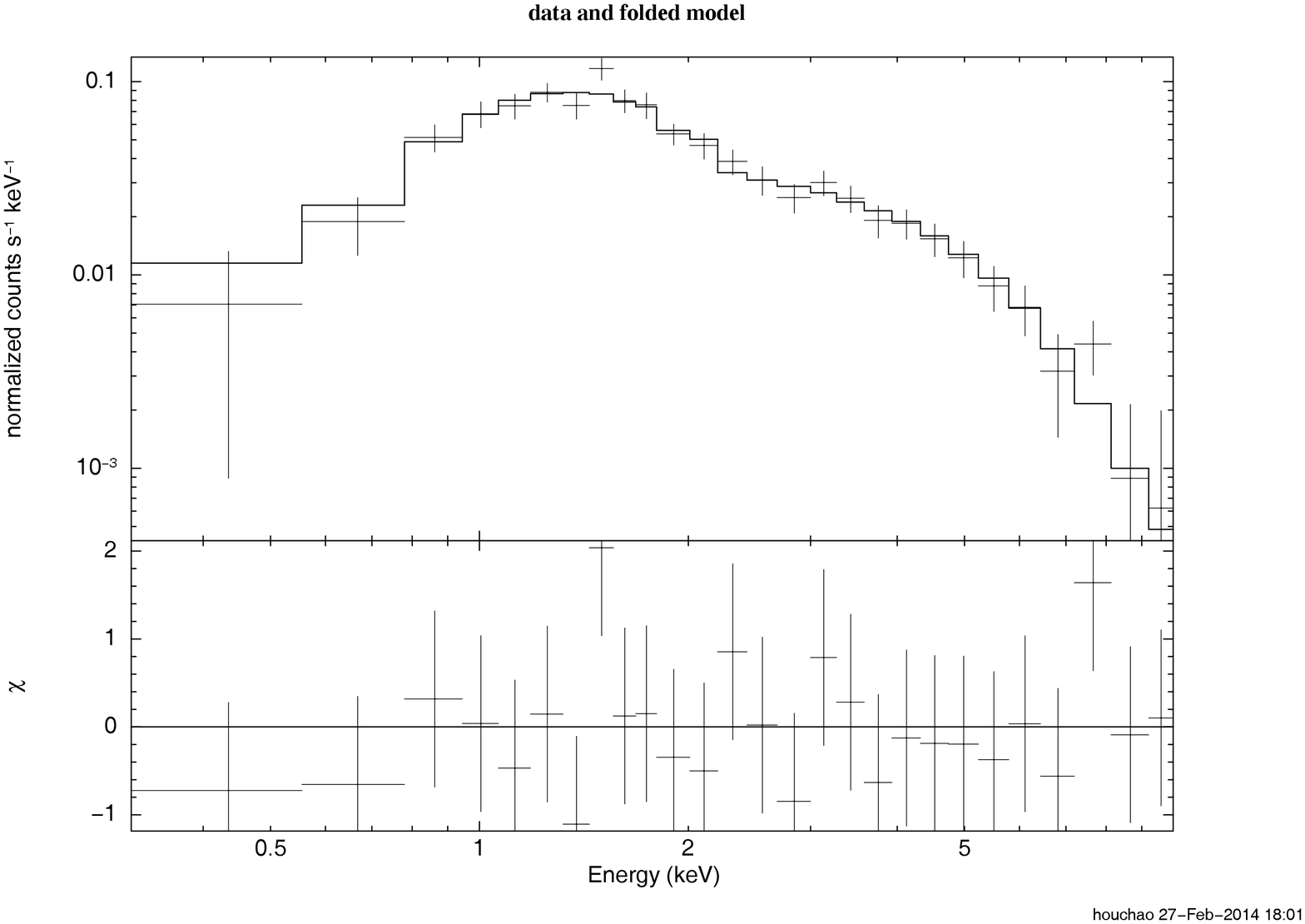}
\figcaption{\label{PWN1-bg2}XMM-Newton MOS2 spectrum for PWN G75.2+0.1.}
\end{center}

\begin{center}
\includegraphics[width=0.5\textwidth]{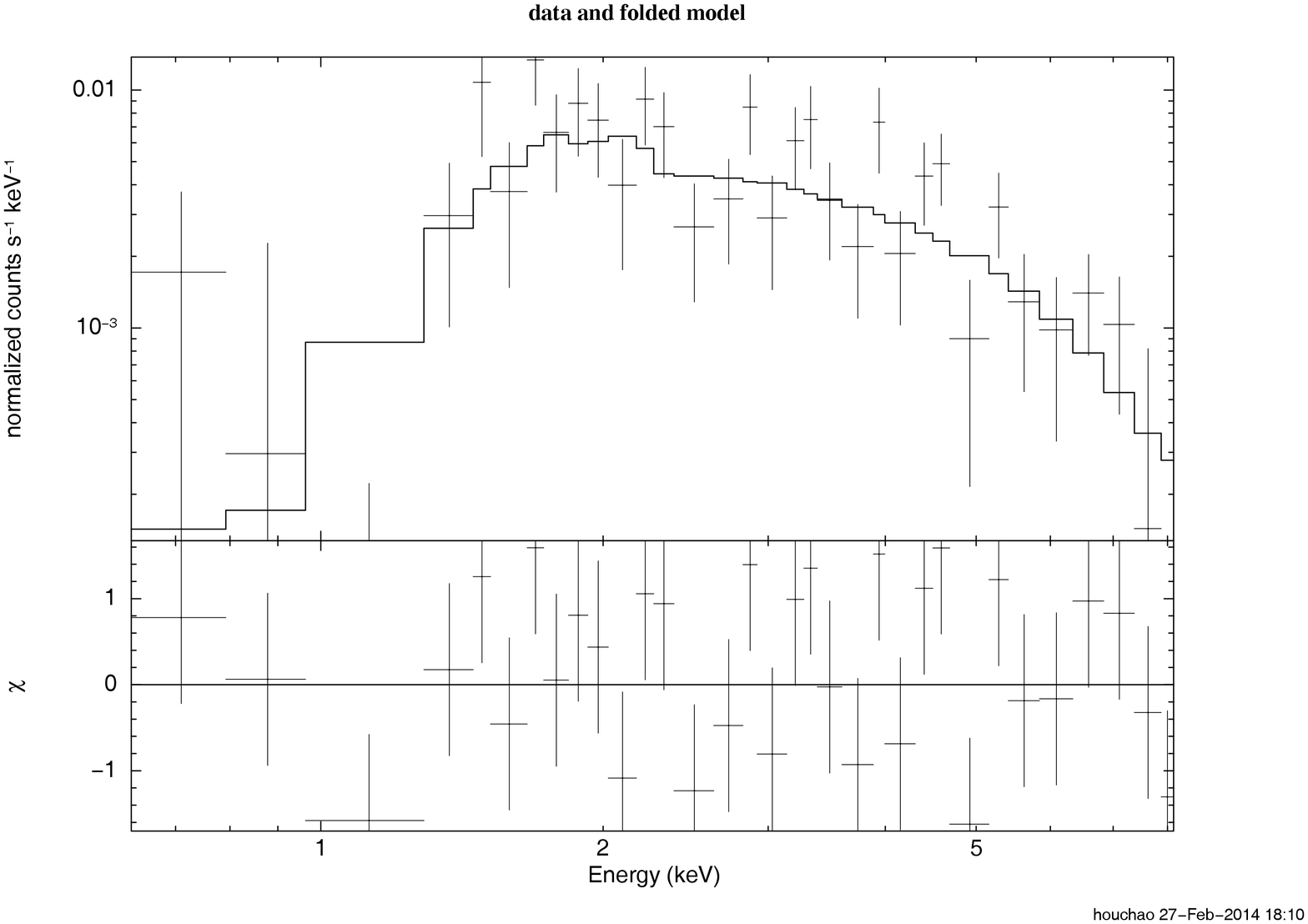}
\figcaption{\label{sh2-104}XMM-Newton pn spectrum for HII region sh2-104.}
\end{center}

The data from both the MOS and pn instruments were analyzed using the
XMM Science Analysis Software (SAS\footnote{http://xmm.esa.int/sas/})
version 13.0.1 with the most recent calibration files. The event files
were created from observation data files (ODFs) using the SAS tasks
epchain and emchain. The events were then filtered to retain only
patterns 0 to 4 for the pn data (0.2$-$15 keV ) and patterns 0 to 12
for the MOS data (0.2$-$12 keV). The data were further filtered to
remove the time intervals of high background rates. The observational IDs
of the data we finally used for spectral analysis are 0404540101 (MOS2
for G75.2+0.1) and 0510011401 (pn for sh2-104). Ancillary response files
and redistribution matrix files are calculated for the corresponding
detector regions. Then we extract the MOS and pn spectra of the
interested source regions.

For PWN G75.2+0.1, we select the extended region with an ellipse
(major radius 6.0' and minor radius 3.7'), to subtract the central pulsar. The background is
extracted from the off-source region near the source. We employ the
X-ray spectral fitting package XSPEC version
12.7.1\footnote{http://heasarc.nasa.gov/xanadu/xspec/} to extract the
source spectrum. An absorbed power-law model is used to fit the data.
The best-fit result and residual of the MOS2 observation of PWN G75.2+0.1
are shown in Fig.~\ref{PWN1-bg2}. For sh2-104, a 0.7' circular area
is selected as the source region. Procedure similar to that for PWN G75.2+0.1
analysis is adopted to extract the spectrum. The results are shown in
Fig.~\ref{sh2-104}. The fitting results are listed in Table
\ref{Fit-results}.

\section{Discussion}

The multi-wavelength observational results of MGRO J2019+37 are shown
in Fig.~\ref{allleptonpwn1}. Here the XMM-Newton results are the
de-absorbed power-law spectrum from PWN G75.2+0.1. According to Milagro
and ARGO-YBJ data, we find that the VHE $\gamma$-ray spectrum peak at
$\sim10$ TeV, which is higher than most of the known TeV emitters
\footnote{http://tevcat.uchicago.edu/}. Up to now (February 2014),
only one of the 147 TeV sources shows similar peak energy as MGRO J2019+37,
i.e. the peak energy of the PWN Vela X  is around 13 TeV \cite{HESSvela}.

VERITAS has resolved the emission of MGRO J2019+37 into a complex
$\gamma$-ray emission region, which is likely composed of multiple sources.
If these sources are independent with each other, we will expect that the
probability to have such a special spectrum with peak energy around
10 TeV is very low. Therefore, it is natural to expect that this complex
$\gamma$-ray emission might have the same origin. MGRO J2019+37 is bright
and extended in TeV band, without counterpart in other wavelengths.
All these properties make it be a mysterious ``dark'' accelerator to
radiate VHE $\gamma$-ray. Understanding its VHE $\gamma$-ray emission
mechanism will be interesting and important.

It is worth noting that the observation periods do not fully overlap for
these detectors. Assuming a distance of 1$-$2 kpc for this source, the
angular extension $\sigma=0.32^{\circ}$ will correspond to a length scale of
5$-$10 pc, which implies a shortest variation time scale of $15-30$ years.
Therefore, we do not expect a significant flux variation over the whole
extended region during the observation time, 2002-2012, of this work.

PWN is a prominent class of VHE $\gamma$-ray sources. G75.2+0.1, PWN of
the pulsar PSR J2021+3651, is a possible candidate to power MGRO J2019+37.
The spin down power is $\dot{E}=3.4\times10^{36}{\rm erg\ s^{-1}}$.
The $\gamma$-ray flux integrated from 1 to 100 TeV based on Milagro
spectrum is $F_{\gamma}=3.2\times10^{-11}{\rm erg\ cm^{-2}s^{-1}}$.
The estimated distance of PSR J2021+3651 ranges from 2 to 12 kpc \cite{fermiJ2021+3651,chandraJ2021+3651}.
The $\gamma$-ray luminosity of this source is:
\begin{equation}
\L_{\gamma}=F_{\gamma}\times(4\pi{d}^{2})=1.5\times10^{34}(d/2\,{\rm kpc})^{2}
{\rm erg\ s^{-1}}.
\end{equation}

The efficiency of VHE $\gamma$-ray emission to the spin down power is
$L_{\gamma}/\dot{E}=0.44\%(d/2\,{\rm kpc})^{2}$. This is consistent
with the range of the $\gamma$-ray efficiency $\eta_{\gamma}=10^{-4}-0.1$
found for other PWNs \cite{pwn2008}.

In the PWN scenario, we could expect a leptonic origin of the multi-wavelength
emission of MGRO J2019+37. The $\gamma$-ray morphology of a PWN might not
be necessarily the same as the X-ray morphology, and the $\gamma$-ray
luminosity can be also much higher than the X-ray luminosity \cite{pwn2009}.
If the supernova explosion occurs in an inhomogeneous medium, the resulting
asymmetric reverse shock will push the pulsar to the direction away from the
higher density medium. In such a scenario, the particles responsible for the
$\gamma$-ray could be the ``relic'' of the PWN, while those responsible
for X-ray could be newly accelerated ones. Therefore, the X-ray image
shows displacement compared with the $\gamma$-ray image. The particle spectrum
to produce X-ray may be also different from that to produce $\gamma$-ray.

It is also possible that the $\gamma$-ray emission is produced by
hadronic cosmic ray interactions, and the X-ray emission is produced by
high energy electrons. In this scenario, there should be more degrees
of freedom of the modeling because there is no direct connection with the
multi-wavelength data. In the following, we will discuss both the leptonic
and hadronic models and explain the multi-wavelength data of MGRO J2019+37.

\subsection{Leptonic model}

A simple leptonic model is constructed to interpret both the X-ray
emission from PWN G75.2+0.1 and $\gamma$-ray emission from MGRO J2019+37.
We assume a uniform distribution of electrons in the vicinity of MGRO J2019+37.
VHE $\gamma$-ray emission is produced by the inverse Compton scattering
of electrons with the interstellar radiation field (ISRF\footnote{We adopt the
local results of the ISRF as an approximation.}), including the cosmic
microwave background, infrared and starlight \cite{porter2005}.
The X-ray emission is produced by the synchrotron radiation of the electrons,
within a confined region surrounding the PWN where the magnetic field
strength is expected to be higher than the average value in the interstellar
medium. A filling factor $f$ is introduced to describe the fraction of
the X-ray emitting volume to the $\gamma$-ray emitting volume.


\begin{center}
\includegraphics[width=0.5\textwidth]{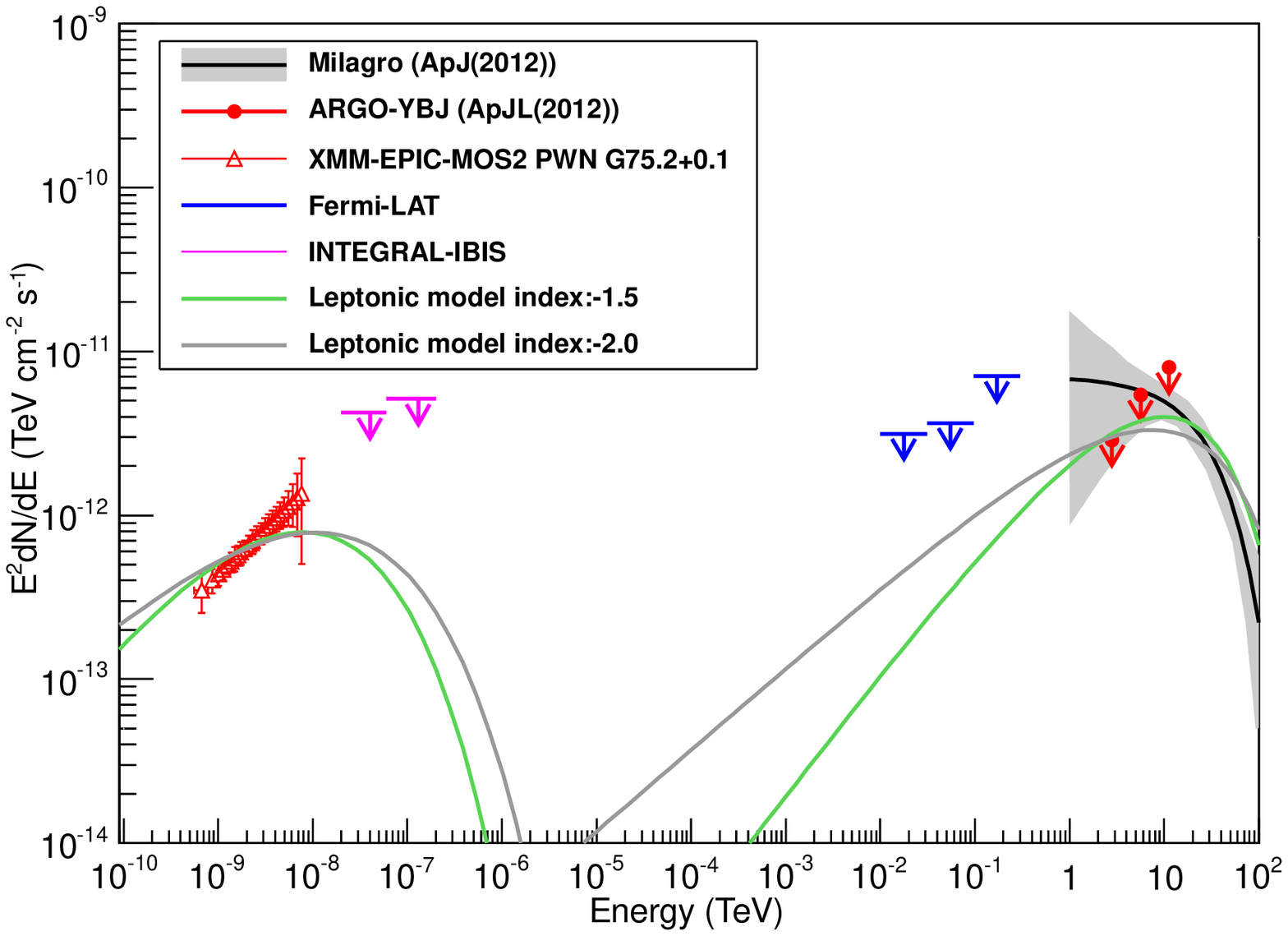}
\figcaption{\label{allleptonpwn1} Multi-wavelength spectrum of MGRO J2019+37.
The shaded region indicates the best-fit spectrum and the $1\sigma$
uncertainty region as measured by Milagro \cite{milagro2012}.
The arrows above 1 TeV show the upper limits derived by the ARGO-YBJ \cite{argocygregion}.
The arrows at keV and GeV show the upper limits derived using data observed by INTEGRAL-IBIS and \emph{Fermi}-LAT, respectively.
The triangles show the flux of PWN G75.2+0.1 using XMM-Newton data.
The solid lines show the two leptonic model expectations as described in the text. (Color online)}
\end{center}

The electron spectrum is assumed to be a power-law function with an
exponential cutoff $dN/dE\propto E^{-\alpha}\exp(-E/E_c)$, where $N$ is the electron number, $E$ is the electron energy, $\alpha$ is spectral index and $E_c$ is cutoff energy. Since most
of the observations give upper limits, we cannot put good constraints on
the model parameters. We choose two values of the electron spectral indices
$\alpha=1.5$ and $2.0$ for illustration. It is possible for a PWN to
give an electron spectrum harder than $2$, e.g., the Crab nebula
\cite{yuan2011}. For $\alpha=1.5$, we have
$E_c\approx50$ TeV, the total energy of electrons above 1 GeV
$W_e(>1\,{\rm GeV})\approx3.2\times10^{46}(d/2\,{\rm kpc})^2$ erg,
the magnetic field to produce X-ray emission of G75.2+0.1 $B\approx30\mu$G,
and the filling factor $f\approx0.3\%$. For $\alpha=2.0$, we have
$E_c\approx90$ TeV, $W_e(>1\,{\rm GeV})\approx8.4\times10^{46}(d/2\,
{\rm kpc})^2$ erg, $B\approx30\mu$G, and $f\approx0.24\%$.
The model expectations are shown in Fig.~\ref{allleptonpwn1}.

The model gives a marginal fit to the data. However, even for $\alpha=1.5$,
the synchrotron spectrum seems softer than the XMM-Newton data of
PWN G75.2+0.1. As shown in Table \ref{Fit-results}, the X-ray spectral
index is about $1.44$, which corresponds to an electron spectral
index $\alpha\approx1.9$. The softening of the expected synchrotron spectrum
should mainly due to the cutoff energy $E_c$. It is possible that, if
PWN G75.2+0.1 is the acceleration source of the high energy electrons,
the electron spectrum will be harder and cutoff at higher energies when
close to the acceleration source. Then the electrons diffuse to the entire
region of MGRO J2019+37, become softer and cutoff earlier. The cooling
time scale for electrons in the ISRF can be estimated as
$\tau_{\rm cool} \approx 3\times10^5(U_{\rm ISRF}/{\rm eV\,cm^{-3}})^{-1}
(E/{\rm TeV})^{-1}$ yr, where $U_{\rm ISRF}$ is the energy density of
the ISRF, and $E$ is the energy of the electrons. Given the lifetime
of PSR J2021+3651 is about $17000$ yr, the cooling energy of electrons
can be as high as tens of TeV, which is consistent with the
required value to fit the data. We can also estimate the diffusion scale
of the electrons. Assuming a diffusion coefficient of $10^{30}$
cm$^2$s$^{-1}$ (approximate value for TeV particles in the Galactic disk
\cite{hess2006}), the diffusion length is estimated to be
$\sim300(D/10^{30}{\rm cm^2s^{-1}})^{0.5}(t/17000\,{\rm yr})^{0.5}$ pc.
Such a value seems too large compared with the spatial extension of
MGRO J2019+37. It seems that the VHE $\gamma$-ray emission should come
from newly accelerated electrons if this scenario works. Detailed modeling
will depend on the time-dependent injection and diffusion of the electrons
which beyond the scope of this work. In this scenario we could expect an
energy dependent $\gamma$-ray morphology to be same as that for HESS J1825$-$137
\cite{aharon06}. Future $\gamma$-ray facilities with higher sensitivity
and angular resolution may test this scenario \cite{chen2013}.

The current spin down power of PSR J2021+3651 is $\dot{E}=3.4\times10^{36}$
erg s$^{-1}$. For an estimated age of about $17000$ years, the time
integrated spin down energy is higher than $1.8\times10^{48}$ erg.
Thus the energy fraction transferred to high energy electrons is then $\eta_e
\approx (2~\rm to~5)\%(d/2\,{\rm kpc})^2$. It shows that PSR J2021+3651 should
be enough to power the $\gamma$-ray emission of MGRO J2019+37.

An alternative scenario is that the electrons are accelerated in a diffuse
region, by e.g. the ensemble of massive OB association in the Cygnus region
\cite{bykov2001}. In this scenario the non-coincidence of the X-ray image
and $\gamma$-ray image can be easily understood. The existence of
acceleration in the extended region of MGRO J2019+37 may also explain
the required hard electron spectrum to reproduce the Milagro measurement
and ARGO-YBJ upper limits. However, the fit to the XMM-Newton spectrum
is not good enough in this scenario. Furthermore, special treatment
of the particle diffusion will be also needed to avoid too large
extension of the source region.

\subsection{Hadronic model}

The $\gamma$-ray emission can be also produced through the decay of neutral
pions which is produced in the inelastic collisions between accelerated cosmic
ray nuclei and the ambient interstellar medium. Still we assume exponential
cutoff power law for the spectrum of the accelerated nuclei (protons for simplicity).
The expected $\gamma$-ray spectra for two illustration values of proton
spectral indices, $\alpha_p=1.5$ and $2.0$, are shown in Fig.~\ref{allhadronic}.
For $\alpha=1.5$ ($2.0$), the adopted cutoff energy is $200$ ($500$) TeV,
and the total energy of protons above 1 GeV is $6.4(16)\times10^{49}$
$(d/2\,{\rm kpc})^2(n/{\rm cm}^{-3})^{-1}$ erg.

\begin{center}
\includegraphics[width=0.5\textwidth]{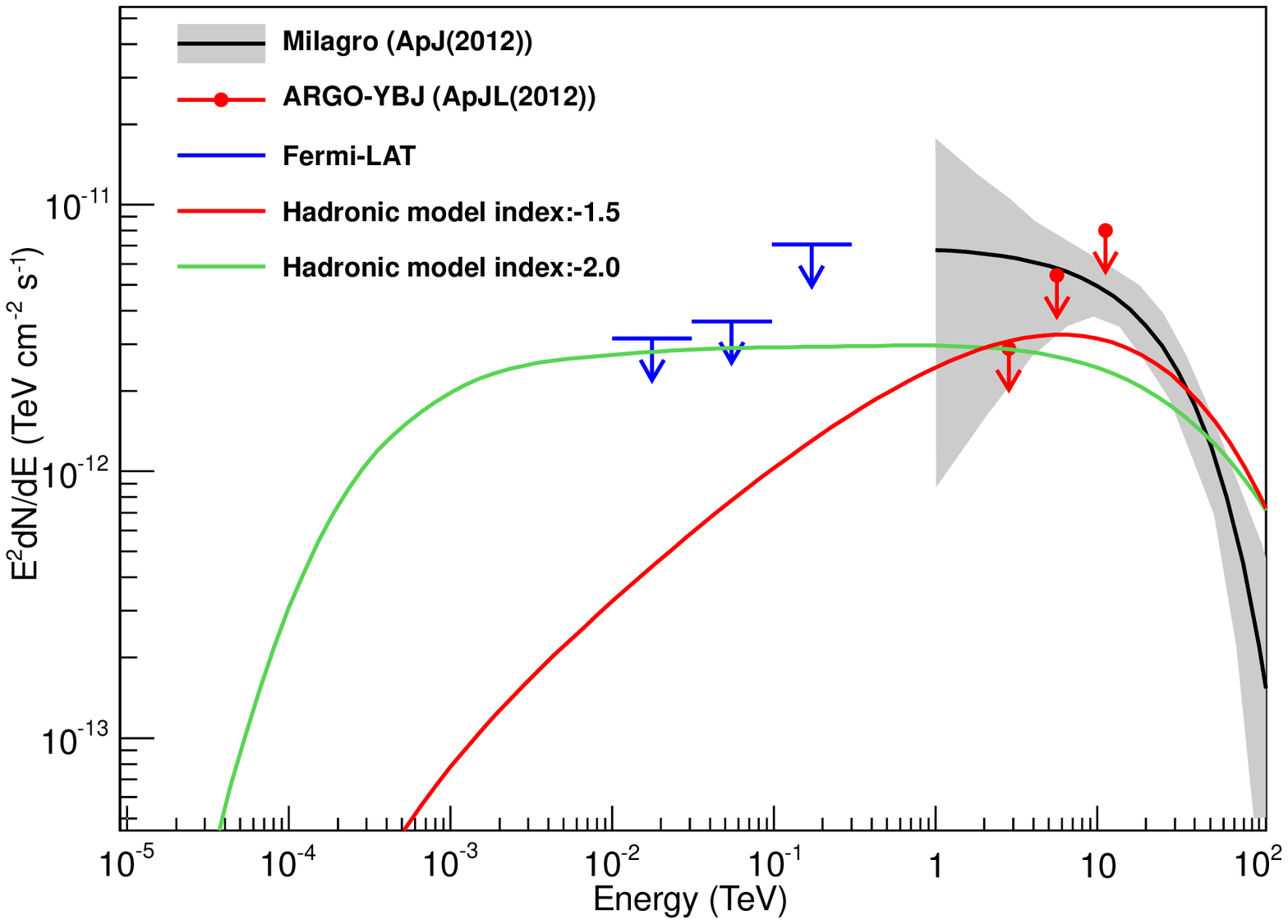}
\figcaption{\label{allhadronic}  Two hadronic model expectations of the $\gamma$-ray spectrum of MGRO
J2019+37, compared with the data. (Color online) }
\end{center}

Fig.~\ref{allhadronic} shows that the hadronic model can
barely fit the $\gamma$-ray data. The exponential cutoff power-law spectrum
of protons seems to be too broad compared with the peak behavior of the VHE
$\gamma$-ray spectrum, unless the proton spectral index is much harder.
The required total energy of protons seems to be higher than the abovely
estimated energy released from spin down of PSR J2021+3651. It is possible
that the total energy release of PSR J2021+3651 is higher than
$1.8\times10^{48}$ erg because the rotation of the pulsar should be
faster in the past. Another possibility is that these protons were
accelerated by the remnant of the supernova which produced PSR J2021+3651.
The total energy of protons seems to be consistent with the canonical
value of $\sim10\%$ of the total kinetic energy released from a typical
supernova explosion, say $10^{51}$ erg. However, the constraints from
particle diffusion as discussed in Sec. 3.1 is also applied here.
A high density concentration or a low diffusion coefficient around
MGRO J2019+37 may be helpful to avoid the too large extension problem
due to diffusion. Similar to the leptonic scenario, the protons can be also
accelerated diffusively in the extended region of MGRO J2019+37.

\section{Summary}

MGRO J2019+37 is a special TeV $\gamma$-ray source in the Northern sky.
This paper presents a collection of the multi-wavelength observations
of MGRO J2019+37 from X-ray to TeV $\gamma$-ray bands. The available
archival data in the direction of MGRO J2019+37 from XMM-Newton at
soft X-ray band, INTEGRAL at hard X-ray band and \emph{Fermi}-LAT at GeV
$\gamma$-ray band are analyzed. There is no corresponding extended signal
in INTEGRAL and \emph{Fermi}-LAT data, and the flux upper limits are obtained.
In XMM-Newton data, emissions from the PWN G75.2+0.1 and the HII region
sh2-104 are found. Spectral analyses of G75.2+0.1 and sh2-104 are performed.
The possible multi-wavelength radiation mechanism of the source is
discussed. It is shown that a leptonic scenario can marginally reproduce
the X-ray to TeV $\gamma$-ray data. The PWN G75.2+0.1 of PSR J2021+3651
might be the acceleration source of the high energy electrons. Although
the hadronic scenario fits the data worse, it should not be excluded
due to the lack of high quality data. The diffuse particle acceleration
from e.g. the ensemble of OB associations can also explain the observational
results.

Due to the lack of detailed observations, no favored model can be obtained
right now. To further unveil the puzzle of the dark
accelerator MGRO J2019+37, observations with more sensitive instruments
at various bands are needed. In the nearly future, the new EAS experiments
such as HAWC, Tibet+MD and LHAASO are expected to be able to achieve a
more accurate observation of $\gamma$-ray sources from 40 GeV to 1 PeV.
The $\gamma$-ray emission mechanism of this source is expected to be
uncovered in the new era of VHE $\gamma$-ray astronomy.

~


\end{multicols}
\vspace{-1mm}
\centerline{\rule{80mm}{0.1pt}}
\vspace{2mm}
\begin{multicols}{2}

\end{multicols}

\clearpage

\end{document}